\documentclass[11pt]{article}
\usepackage{blois,epsfig,graphicx}
\usepackage[latin1]{inputenc}
\usepackage[OT1]{fontenc}
\usepackage{epstopdf}

\def\be{\begin{equation}}
\def\ee{\end{equation}}
\def\bea{\begin{eqnarray}}
\def\eea{\end{eqnarray}}

\newcommand{\vect}[1]{\mathbf{#1}}   
\begin{document}
\vspace*{4cm}
\title{UHECRs from the Radio Lobes of AGNs}

\author{ F. Fraschetti$^{1,2,~}$\footnote{Corresponding author: federico.fraschetti@cea.fr}, and F. Melia$^3$ }

\address{$^1$ Laboratoire AIM, CEA/DSM - CNRS - Univ. Paris Diderot,\\
Irfu/SAp, F-91191 Gif sur Yvette C\'edex, France;\\ $^2$ 
LUTh, Observatoire de Paris, CNRS-UMR8102 and Universit\'e Paris VII,\\
5 Place Jules Janssen, F-92195 Meudon C\'edex, France;\\
$^3$ Department of Physics and
Steward Observatory, The University of Arizona, Tucson, AZ 85721,
USA}

\maketitle\abstracts{We report a stochastic mechanism of particle acceleration 
from first principles in an environment having properties like those
of Radio Lobes in AGNs. We show that energies $\sim 10^{20}$ eV are reached in $\sim 
10^6$ years for protons. Our results reopen the question regarding the nature of 
the high-energy cutoff in the observed spectrum: whether it is due solely to propagation 
effects, or whether it is also affected by the maximum energy permitted by the 
acceleration process itself.}

\section{Introduction}

The search for the origin of the UHECRs still represents one the major 
challenges of theoreti\-cal astrophysics. Theoretical models may be divided 
into two classes: the so-called ``bottom-up''scenarios~\cite{ta04}, in which 
a specific process of acceleration in a particular astrophysical object leads 
to UHEs; and the so-called ``top-down'' prescription~\cite{bs00}, in which UHE 
particles are produced through the decay of super-heavy dark matter particles 
or by collision among cosmic strings or by topological defects. The recent 
measurement by Auger, demonstrating a low fraction of high-energy photons in 
the CR distribution, rule out the top-down models, in which 
the UHECRs represent the decay products of high-mass particles created in the 
early Universe~\cite{semikoz07}. The top-down models based on topological defects, 
however, are still compatible with the current data and could be constrained by 
future experiments.

Recently a steepening in the UHECR spectrum has been reported by both the 
HiRes~\cite{H07} and Auger~\cite{auger2007} collaborations. This result 
may be a strong confirmation of the predicted Greisen-Zatsepin-Kuzmin (GZK) 
cutoff due to photomeson interactions between the UHECRs and low-energy 
photons in the cosmic microwave background (CMB) radiation~\cite{greisen66,zatsepin66}.

The most telling indicator for the possible origin of these UHECRs is
the discovery by Auger of their clustering towards nearby ($\sim 75$
Mpc) AGNs along the supergalactic plane. However, the question remains 
open regarding the mechanism of acceleration to such high energies and 
on the origin of the observed cutoff in the spectrum, i.e., if it is due 
solely to the GZK effect, or whether it also points to an intrinsic limit 
to the acceleration efficiency.

UHECRs generation scenarios include the so-called first-order Fermi
acceleration in GRBs, Pulsar Wind Bubbles, and also relativistic second order 
Fermi acceleration~\cite{f49,g08}. We report here a treatment of particle acceleration
in the lobes of radio-bright AGNs from first principles~\cite{fm08}, considering the acceleration 
of charged particles via random scatterings (a second-order process) with 
fluctuations in a turbulent magnetic field.

\section{Model of magnetic turbulence}

In our treatment, we follow the three-dimensional motion of {\it individual} 
particles within a time-varying field. By avoiding the use of equations 
describing statistical averages through the phase space distribution function 
of a given population of particles, we mitigate our dependence on unknown 
factors, such as the diffusion coefficient.  We also avoid the need to
use the Parker approximation~\cite{padma} in the transport equation. 
The remaining unknowns are the energy partition between 
turbulent and background fields, and the turbulent spectral distribution, 
though this may reasonably be assumed to be Kolmogorov. For simplicity, 
we assume that the magnetic energy is divided equally between the 
two components; the actual value of this fraction does not produce 
any significant qualitative differences in our results. 

We calculate the trajectory of a test particle with charge $e$ and mass
$m$ in a magnetic field $\vect{B}(t, \vect{r}) =mc\vect{\Omega}(t, \vect{r})/e$, 
where $c$ is the speed of light in vacuum. The particle motion is obtained as 
a solution of the Lorentz equation
\begin{equation}
\frac{d\vect{u}(t)}{dt} = \delta \vect{\cal E}(t, \vect{r}) + \frac{\vect{u}(t) 
\times\vect{\Omega}(t,\vect{r})}{\gamma(t)}\;,
\label{lorentz}
\end{equation}
where $\vect{u}$ is the three-space vector of the four-velocity $u^\mu
=\left(\gamma, \gamma {\vect{v}}/{c}  \right)$, $t$ is the time in 
the rest frame of the source, and $\gamma$ is the Lorentz
factor $\gamma = 1/ \sqrt{1-(v/c)^2}$. The quantity $\vect{\Omega}$
in equation (\ref{lorentz}) is given by $\vect{\Omega}(t, \vect{r}) = \vect{\Omega}_0
+\delta \vect{\Omega}(t, \vect{r})$, where $\vect{\Omega}_0 =e \vect{B}_0/mc$
and $\vect{B}_0$ is the background magnetic field. The time variation of the 
magnetic field, however, induces an electric field $\delta \vect{\cal{E}}(t, 
\vect{r}) = (e/mc) \vect{E}(t, \vect{r})$ according to Faraday's law. 
We ignore any large-scale background electric fields; this is a reasonable 
assumption given that currents would quench any such fields within the 
radio lobes of AGNs. 

We follow the Giacalone-Jokipii~\cite{gj94} prescription for generating the
turbulent magnetic field, including a time-dependent phase factor to allow
for temporal variations. This procedure calls for the random generation 
of a given number $N$ of transverse waves $\vect{k}$ at every point of physical 
space where the particle is found, each with a random direction defined by angles 
$\theta(k_i)$ and $\phi(k_i)$. This form of the fluctuation satisfies $\nabla \cdot 
\vect{B}=0$. We write 
\begin{equation}
\delta \vect{\Omega}(t, \vect{r}) = \sum_{i=1}^N\Omega(k_i) [\cos\alpha(k_i)
\hat{\bf y}' \pm i\sin\alpha(k_i)\hat{\bf z}']e^{\left[i(k_i x' - \omega_i t +
\beta(k_i))\right]}\;.
\label{omega}
\end{equation}
The primed reference system $(x',y',z')$ is related 
to the lab-frame coordinates $(x,y,z)$ via a rotation in terms of $\theta(k_i)$
and $\phi(k_i)$. For each $k_i$, there are 
5 random numbers: $0<\theta(k_i)<\pi$, $0<\phi(k_i)<2\pi$, $0<\alpha(k_i)<2\pi$,
$0<\beta(k_i)<2\pi$ and the sign $\pm$ indicating the sense of polarization.
We use the dispersion relation for transverse non-relativisitc Alfven waves in the background plasma: 
$\omega(k_i) = v_A k_i cos\theta(k_i)$, where $v_A =  B_0/\sqrt{4\pi m_p n}$ is the non relativistic
Alfven velocity in a medium with background magnetic field $B_0$ 
and number density $n$, being $m_p$ the proton mass, and $\theta(k_i)$ is
the angle between the wavevector $k_i$ and $B_0$.
The background plasma is assumed
to have a background number density $n \sim 10^{-4}$ cm$^{-3}$, a reasonable value 
for the radio lobes of AGNs.

The amplitudes of the magnetic fluctuations are assumed to be generated by 
Kolmogorov turbulence, so
\begin{equation}
\Omega(k_i) = \Omega(k_{min}) \left(\frac{k_i}{k_{min}}\right)^{-\Gamma/2}\;,
\end{equation}
where $k_{min}$ corresponds to the longest wavelength of the fluctuations and 
$\Gamma = 5/3$. Finally, the quantity 
$\Omega(k_{min})$ is computed by requiring that the energy density of the magnetic 
fluctuations equals that of the background magnetic field: ${B_0}^2 / 8\pi$.

We choose N=2400 values of $k$ evenly spaced on a logarithmic scale; considering that the
turbulence wavenumber $k$ is related to the turbulent length scale $l$ by $k = 2\pi / l$,
we adopt a range of lengthscales from $l_{min} = 10^{-1}\,v_0/\Omega_0$ to $l_{max} = 
10^{9}\,v_0/\Omega_0$, where $v_0$ is the initial velocity of the 
particle and $\Omega_0$ is its gyrofrequency in the background magnetic field.
Thus the dynamic range covered by $k$ is $k_{max} / k_{min} = l_{max} / l_{min} = 10^{10}$ 
and the interaction of particle with the turbulent waves is gyroresonant at all times.
The particles passing through this region are released at a
random position inside the acceleration zone, which for simplicity is chosen to
be a sphere of radius ${\mathcal R}$, with a fixed initial velocity $u_0$ pointed
in a random direction. The initial value of the Lorentz factor $\gamma_0 = \sqrt{1
+ {u_0}^2} = 1.015$ is chosen to avoid having to deal with ionization losses
for the protons and ions. 

Assuming that both the radio and CMB intensity fields are isotropic, we take these 
energy losses into account using the following angle-integrated power-loss rate: 
\begin{equation}
-\frac{dE}{dt} = \frac{4}{3} \sigma_T(m) c \gamma^2
\left(\frac{B^2}{8\pi} + U_{R} + U_{CMB}\right)\;,
\label{loss}
\end{equation}
where $\sigma_T(m) = 6.6524 \times (m_e/m)^2\, 10^{-25}$ cm$^2$ is the Thomson 
cross section for a particle of mass $m$,
$B^2/(8\pi) = (2 {B_0}^2)/(8\pi)$ is the total energy density of the
magnetic field, and $U_{R}$ is the photon energy density inside a typical Radio
Lobe, for which we assume a standard luminosity density corresponding to
the Fanaroff-Riley class II of galaxies (with a luminosity $L = 5\times10^{25}$
W Hz$^{-1}$ sr$^{-1}$ at $178$ MHz), and a radius ${\mathcal R} = 30$ kpc, the
size of our spherical acceleration zone. For the CMB, we use $U_{CMB} = a T^4 
= 4.2 \times 10^{-13}$ erg cm$^{-3}$. 

In a region where magnetic turbulence
is absent or static, a given test particle propagates by ``bouncing" randomly
off the inhomogeneities in $\vect{B}$, but its energy remains constant. The
field we are modeling here, however, is comprised of transverse plane waves
(see equation \ref{omega}), and collisions between the test particle and these
waves produces (on balance) a net acceleration as viewed in the lab frame.

\begin{figure}
\psfig{figure=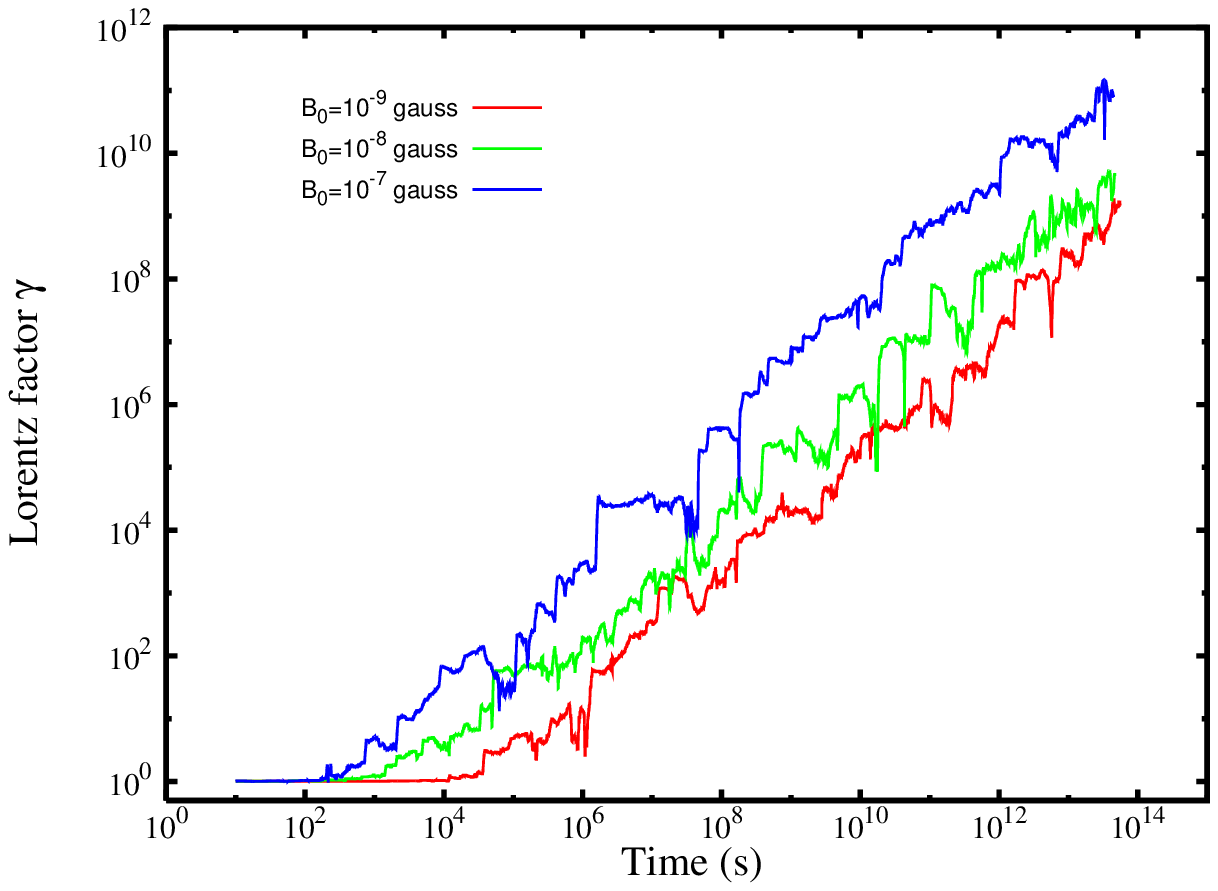,height=2.2in}
\psfig{figure=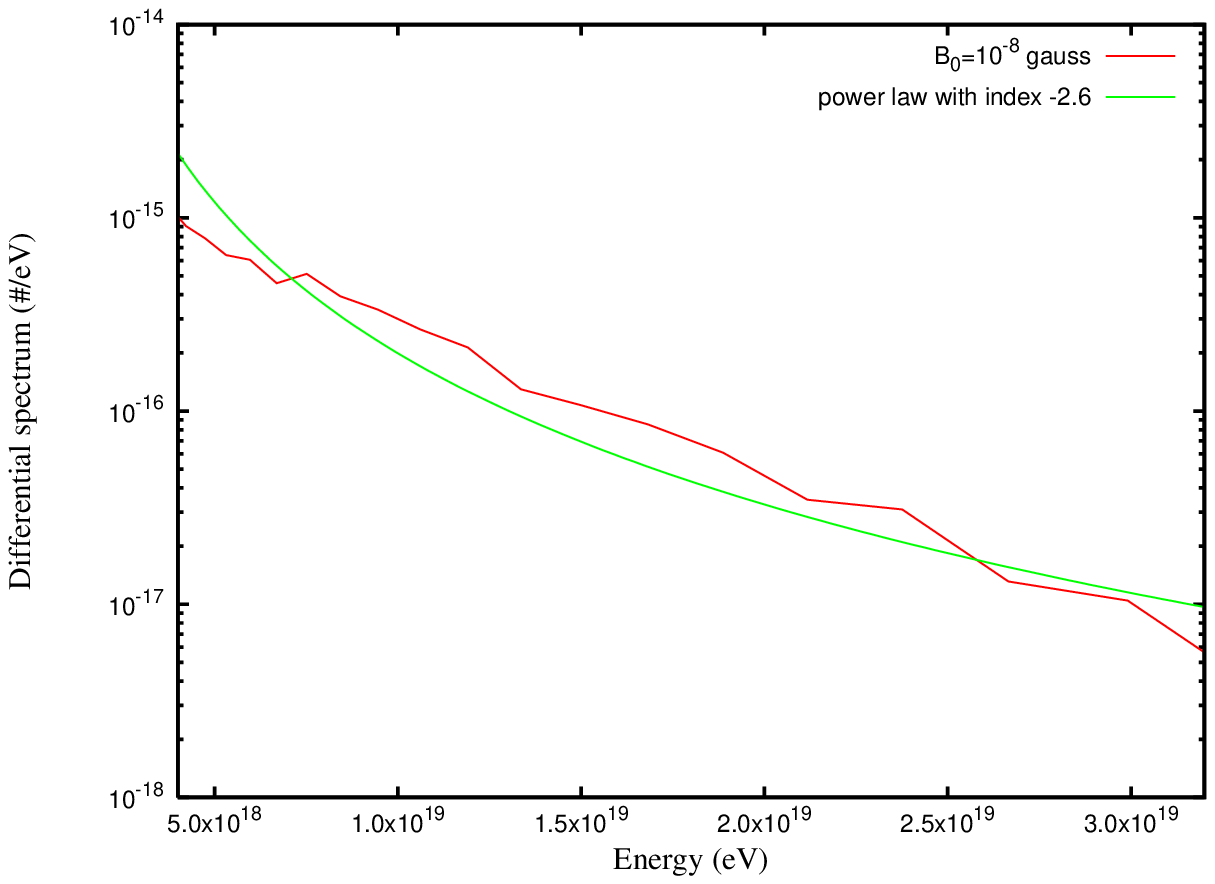,height=2.2in}
\caption{{\it Left}: Simulated time evolution of the Lorentz factor $\gamma$
for a proton propagating through a time-varying turbulent magnetic field.
The particle is followed until it leaves the acceleration zone and enters
the intergalactic medium. The acceleration timescale $\Delta t$ is inversely 
proportional to the background field $B_0$. Therefore, as expected, a larger
$B_0$ produces a more efficient acceleration. {\it Right}: Calculated differential spectrum for 500 protons in the energy range $\log(E/eV) = [18.6-19.5]$ in a background magnetic field of $B_0 = 10^{-8}$ gauss. 
For comparison a power law with index $2.6$ is shown.
\label{gamma1}}
\end{figure}

In Figure \ref{gamma1} ({\it left}), we plot the time evolution of the particle Lorentz factor $\gamma$ for
three representative values of the background field $B_0$: $10^{-7}$, $10^{-8}$, and
$10^{-9}$ gauss. We see the particle undergoing various phases of acceleration and
deceleration as it encounters fluctuations in $\vect{B}$. In Figure \ref{gamma1} ({\it right}), we compare
a differential injection spectrum for a population of 500 protons for energy $E > 4\times 10^{18}$ eV.
The observed spectrum may be affected by the cosmological evolution in source density.
However, a likelihood analysis~\cite{gks07} of the dependence of the observed distribution
on input parameters has already shown that, in the case of pure proton-fluxes of primaries,
for $\alpha \sim 0$, where $\alpha$ is the evolution index in the source density, 
the HiRes observations are compatible with a power-law injection spectrum with index $-2.6$. 

From our sampling of the various physical parameters, we infer 
that $B_0$ should lie in the range $(0.5,5) \times 10^{-8}$ gauss 
in order to produce UHECRs with the observed distribution. We note, however, 
that the particle distribution calculated for energies above 50 EeV does not 
include the GZK effect, which becomes progressively more important as the 
energy approaches $10^{20}$ eV.

\section{Conclusion}

In view of the very good match between our theoretical simulation and the Auger
observations, it is worth emphasizing that this calculation was carried out
without the use of several unknown factors often required in approaches involving
a hybrid Boltzmann equation to obtain the phase-space particle distribution. 
In addition, we point out that the acceleration mechanism we have invoked
here is sustained over 10 orders of magnitude in particle energy, 
and the UHECRs therefore emerge naturally---without the introduction
of any additional exotic physics---from the physical conditions thought to be
prevalent within AGN giant radio lobes.

As the Auger observatory gathers more data and improves the statistics, our
UHECR source identification will continue to get better. Eventually, we should
be able to tell how significant the GZK effect really is, and whether the cutoff
in the CR distribution is indeed due to propagation effects, or whether it is
primarily the result of limitations in the acceleration itself. Given the fact
that energies as high as $\sim 10^{20}$ eV may be reached within typical 
radio lobes, it is possible that both of these factors must be considered in
future refinements of this work.

\section*{Acknowledgments}
FF wishes to thank the Organizers of the Conference. The work of FF was supported by CNES (French Space Agency) and was carried out
at CEA/Saclay and partially at the Center for Particle Astrophysics
and Cosmology (APC) in Paris.

\end{document}